\documentclass[preprintnumbers,amsmath,amssymb,eqsecnum]{revtex4}

\usepackage{epsf}
\usepackage{graphicx}  
\usepackage{dcolumn}   
\usepackage{bm}        

\begin{document}

\preprint{hep-th/06mmnnn}

\title{Entropy Function for 4-Charge Extremal Black Holes in Type IIA Superstring Theory}

 \author{Rong-Gen Cai\footnote{e-mail address:
cairg@itp.ac.cn}}

\address{
  Institute of Theoretical Physics, Chinese
Academy of Sciences,
 P.O. Box 2735, Beijing 100080, China}

\author{Da-Wei Pang\footnote{e-mail address:
pangdw@itp.ac.cn}}
\address{Institute of Theoretical Physics, Chinese
Academy of Sciences,
 P.O. Box 2735, Beijing 100080, China\\
 and Graduate University of the Chinese Academy of Sciences}

\begin{abstract}
We calculate the entropy of 4-charge extremal black holes in Type
IIA supersting theory by using Sen's entropy function method.
Using the low energy effective actions in both $10D$ and $4D$, we
find precise agreements with the Bekenstein-Hawking entropy of the
black hole. We also calculate the higher order corrections to the
entropy and find that they depend on the exact form of the higher
order corrections to the effective action.
\end{abstract}
\maketitle

\section{Introduction}
D-branes play a very important role in string theory because
several kinds of extremal black holes can be constructed out of
various D-brane configurations and the Bekenstein-Hawking entropy
for such kinds of black holes can be explained  by counting the
degeneracy of the microstates of such configurations, see, for
example, \cite{count}. It is also known that for black holes in
higher derivative gravity, the Bekenstein-Hawking entropy formula
$S=\frac{1}{4}A$ does not hold any more. However, Wald  has
presented a method for calculating entropy of black holes in any
diffeomorphism invariant gravity~\cite{wald,jm,dirty}, which can
be used to compute entropy of black holes in higher derivative
gravity.

Recently, based on the Wald's method, Sen has shown that for a
certain class of extremal black holes, the entropy is given by a
so-called ``entropy function" at the extremum~\cite{sen}. The steps
are given as follows:

i) Suppose the near horizon geometry of a D-dimensional extremal
black hole  has the form $AdS_{2}\times S^{D-2}$. The part of
$AdS_2$ has the form $-r^{2}dt^{2}+dr^{2}/r^{2}$. The metric of near
horizon geometry of the black hole is parametrized by two constants
$v_1$ and $v_2$, which stand for the sizes of $AdS_2$ and $S^{D-2}$.

ii) Assume the black hole configuration $AdS_{2}\times S^{D-2}$ is
supported by the electric and magnetic fields and the constant
values $u_i$ of various scalar fields. Define an entropy function
by carrying out the integration of the Lagrangian density of the
gravity theory over $S^{D-2}$ enclosing the black hole and then
making a Legendre transform with respect to the electric fields.
The entropy function is a function of $v_1$ and $v_2$, the scalar
fields $u_{i}$, the electric charges $q_{i}$ and the magnetic
charges $p_{i}$.

iii)  For given  $q_i$ and $p_i$,   $v_1$, $v_2$ and $u_{i}$ can
be determined by extremizing the entropy function with respect to
 $v_{1}$, $v_{2}$ and $u_i$ themselves. Furthermore the entropy
 of the black hole is given by the value of the entropy function at the extremum
point by substituting $v_1$, $v_2$ and $u_i$ into back the entropy
function.

This is a very simple and useful method for calculating the entropy
of such kinds of black holes, especially one can easily find the
corrections to the entropy due to the higher derivative terms in the
effective action. Some related works see~\cite{PP}~--~\cite{Af}.

Some extremal black holes in type II superstring theory have
$AdS_{3}$ as part of their near horizon geometries in ten
dimensions, instead of $AdS_2$. It turns out that the Sen's
entropy method can still be used in that case. The case of D1-D5-P
black holes in type IIB superstring theory has been discussed very
recently in~\cite{Ir}. In the present paper we will discuss
D2-D6-NS5-P black holes in type IIA superstring theory.

The rest of the paper is organized as follows: In Sec.~II we will
briefly review some properties of D2-D6-NS5-P black holes as
solutions of type IIA supergravity. We apply Sen's method to
D2-D6-NS5-P black holes in ten dimensions in Sec.~III. Then after
doing dimensional reduction on the effective action to four
dimensions, we calculate the entropy function in four dimensions
in Sec.~IV. Finally we compute the $\alpha'^3$ corrections to the
entropy function in Sec.~V. We summarize and discuss our main
results in the last section.

\section{4-charge black hole in type IIA superstring theory}

Let us review some properties of 4-charge black hole in type IIA
supersting theory~\cite{maphd}. Type IIA supergravity consists of
two sectors. One is the (NS,NS) sector containing the metric
$g_{\mu\nu}$, a two form $B_{\mu\nu}$ and the dilaton $\phi$. The
other is the (R,R) sector, which has a one form $A_{\mu}$ and a
three form $C_{\mu\nu\rho}$. The bosonic part of the effective
action in string frame can be written as
\begin{eqnarray}
\label{2eq1}
 S&=&\frac{1}{16\pi G^{10}_{N}}\int d^{10}x\sqrt{-g}
 [e^{-2\phi}(R+4(\nabla\phi)^{2}-\frac{1}{3}H^{2})
 -G^{2}-\frac{1}{12}F'^{2}\nonumber \\
\nonumber \\
 & &-\frac{1}{288}\epsilon^{\mu_{1}\cdots\mu_{10}}F_{\mu_{1}
 \mu_{2}\mu_{3}\mu_{4}}F_{\mu_{5}\mu_{6}\mu_{7}\mu_{8}}B_{\mu_{9}\mu_{10}}],
\end{eqnarray}
where $G^{10}_N$ is the Newtonian constant in ten dimensions,
$G=dA$, $H=dB$, $F=dC$ and $F'=dC+2A\wedge H$ are the field
strengths associated with each of the differential forms.

The $4D$ extremal black hole with finite horizon area can be
constructed by wrapping $D6$ branes on $T^{6}=T^{4}\times
S'_{1}\times S_{1}$, $D2$ branes on $S'_{1}\times S_{1}$
(directions 4, 9), $NS5$ branes on $T^{4}\times S_{1}$ (directions
5, 6, 7, 8, 9) and momentum flowing along $S_{1}$ (direction 9).
Note that the $NS5$ brane does not break any additional
supersymmetry and the final configuration still preserves $1/8$ of
the original supersymmetries. The $4D$ extremal black hole,
constructed this way, written in $10D$ string frame, has the form
\begin{eqnarray}
ds^{2}_{str}&=&f^{-\frac{1}{2}}_{2}f^{-\frac{1}{2}}_{6}(-dt^{2}+dx_{9}^{2}+k(dt-dx_{9})^2)
         \nonumber\\
            & &+f_{s5}f^{-\frac{1}{2}}_{2}f^{-\frac{1}{2}}_{6}dx_{4}^{2}+f^{\frac{1}{2}}_{2}
            f^{-\frac{1}{2}}_{6}(dx_{5}^{2}+\cdots dx_{8}^{2})\nonumber\\
            & &+f_{s5}f^{\frac{1}{2}}_{2}f^{\frac{1}{2}}_{6}(dx_{1}^{2}+\cdots
            dx_{3}^{2}),\nonumber\\
& &e^{-2\phi}=f_{s5}^{-1}f_{2}^{-1/2}f_{6}^{3/2}, \quad H_{ij4}=\frac{1}{2}\epsilon_{ijk}
\partial_{k}f_{s5}\quad i,j,k=1,2,3,\nonumber\\
& &C_{049}=\frac{1}{2}(f_{2}^{-1}-1),\quad(dA)_{ij}=\frac{1}{2}\epsilon_{ijk}\partial_{k}f_{6}\quad
i,j,k=1,2,3,
\end{eqnarray}
where $\epsilon_{ijk}$ is the flat space epsilon tensor. The
harmonic functions are
\begin{equation}
\begin{array}{ll}
f_{2}=1+\frac{Q_{2}}{r}, \quad
f_{s5}=1+\frac{Q_{5}}{r},\\
\\
f_{6}=1+\frac{Q_{6}}{r},\quad k=\frac{Q_{P}}{r},
\end{array}
\end{equation}
where $Q_{2}=c^{(4)}_{2}N_{2}, Q_{5}=c^{(4)}_{5}N_{5},
Q_{6}=c^{(4)}_{6}N_{6}$ and $Q_{P}=c^{(4)}_{P}N_{P}$ and the
coefficients $c^{(4)}$'s are
\begin{equation}
\begin{array}{ll}
c^{(4)}_{2}=\frac{4G^{4}_{N}R_{4}R_{9}}{g_{s}\alpha'^{\frac{3}{2}}},
\quad c^{(4)}_{s5}=\frac{\alpha'}{2R_{4}},\\
c^{(4)}_{6}=\frac{g_{s}\alpha'^{\frac{1}{2}}}{2},\quad
c^{(4)}_{P}=\frac{4G^{4}_{N}}{R_{9}}
\end{array}
\end{equation}
and $N_{2}, N_{5}, N_{6}$ and $N_{P}$ are integers. $G^{4}_{N}$
denotes the $4D$ Newtonian constant while $R_{4}$ and $R_{9}$ are
the radii of $S'_{1}$ and $S_{1}$.

We can obtain the $4D$ metric in string frame by the standard
dimensional reduction
\begin{equation}
ds_{4}^{2}=-f^{-\frac{1}{2}}_{2}f^{-\frac{1}{2}}_{6}(1+k)^{-1}dt^{2}
+f_{s5}f^{\frac{1}{2}}_{2}f^{\frac{1}{2}}_{6}(dr^{2}+r^{2}d\Omega_{2}^{2}).
\end{equation}
This describes a $4D$ black hole and its horizon is located at
$r=0$.  The near horizon ($r \to 0$) geometry of the 4-charge
extremal black hole is given by
\begin{equation}
ds_{near}^{2(4)}=-\frac{r^{2}}{Q_{P}\sqrt{Q_{2}Q_{6}}}dt^{2}
+\frac{Q_{5}\sqrt{Q_{2}Q_{6}}}{r^{2}}dr^{2}+Q_{5}\sqrt{Q_{2}Q_{6}}(d\theta^{2}+\sin^{2}\theta
d\phi^{2}),
\end{equation}
clearly it is of the form $AdS_{2}\times S^{2}$. The near-horizon
geometry of the black hole in $10D$ dimensions has the metric
\begin{eqnarray}
ds^{2}&=&(\frac{Q_{P}-r}{\sqrt{Q_{2}Q_{6}}}dt^{2}-2\frac{Q_{P}}{\sqrt{Q_{2}Q_{6}}}dtdx_{9}
+\frac{Q_{P}+r}{\sqrt{Q_{2}Q_{6}}}dx_{9}^{2}
+\frac{Q_{5}\sqrt{Q_{2}Q_{6}}}{r^{2}}dr^{2})\nonumber\\
      & &+Q_{5}\sqrt{Q_{2}Q_{6}}(d\theta^{2}+\sin^{2} \theta
d\phi^{2}) +\frac{Q_{5}}{\sqrt{Q_{2}Q_{6}}}dx_{4}^{2}
+\sqrt{\frac{Q_{2}}{Q_{6}}}(dx_{5}^{2}+\cdots dx_{8}^{2}).
\end{eqnarray}
It is of the form $AdS_3 \times S^2 \times S^1 \times T^4$.
 The Bekenstein-Hawking entropy of the black hole is
\begin{equation}
S_{BH}=\frac{A_{4}}{4G^{4}_{N}}=2\pi\sqrt{N_{2}N_{5}N_{6}N_{P}} ,
\end{equation}
The statistical entropy can be derived by counting the degree of
freedom of the corresponding D-brane configurations. The result is
in precise agreement with the Bekenstein-Hawking
entropy~\cite{maphd}.


\section{Entropy function of 4-charge black hole in $D=10$ }

We start from the ten dimensional string frame metric and write
down the near horizon field configuration
\begin{eqnarray}
\label{3eq1}
ds^{2}&=&v_{1}(\frac{Q_{P}-r}{\sqrt{Q_{2}Q_{6}}}dt^{2}-2\frac{Q_{P}}{\sqrt{Q_{2}Q_{6}}}dtdx_{9}
+\frac{Q_{P}+r}{\sqrt{Q_{2}Q_{6}}}dx_{9}^{2}
+\frac{Q_{5}\sqrt{Q_{2}Q_{6}}}{r^{2}}dr^{2})\nonumber\\
\nonumber\\
      & &+v_{2}Q_{5}\sqrt{Q_{2}Q_{6}}(d\theta^{2}+\sin^{2} \theta
d\phi^{2}) +v_{3}\frac{Q_{5}}{\sqrt{Q_{2}Q_{6}}}dx_{4}^{2}
+v_{4}\sqrt{\frac{Q_{2}}{Q_{6}}}(dx_{5}^{2}+\cdots
dx_{8}^{2})\nonumber\\
\nonumber\\
& &H_{\theta\phi 4}\equiv p_{1}\sin\theta=-\frac{Q_{5}}{2}\sin\theta, \quad G_{\theta\phi}\equiv p_{2}\sin\theta=-\frac{Q_{6}}{2}\sin\theta,\nonumber\\
\nonumber\\
& &F_{t49r}=e_{1}, \quad e^{-2\phi}=u_{s}.
\end{eqnarray}
Note that here the parameters $v_{3}$ and $v_{4}$, which describe
the sizes of $S_{1}'$ and $T^{4}$, have been introduced, except
for $v_1$ and $v_2$. This is different from the consideration
in~\cite{Ir} for 3-charge black hole in Type IIB supergravity. In
addition,  the D2 branes are considered here as electric field
sources while D6 and NS5 branes as magnetic field sources.

 The general form of Wald formula for computing black hole entropy is~\cite{dirty}
\begin{equation}
S_{BH}=4\pi\int_{H}dx_{H}\sqrt{g_{H}}\frac{\partial\mathcal{L}}{\partial
R_{\mu\nu\lambda\rho}}g_{\mu\lambda}^{\bot}g_{\nu\rho}^{\bot},
\label{pp}
\end{equation}
where ${\cal L}$ is the Lagrangian density of the gravity theory
under consideration, $g_{H}$ is the determinant of the horizon
metric and $g_{\mu\nu}^{\bot}$ denotes the orthogonal metric
obtained by projecting onto subspace orthogonal to the horizon.
For the general form of the metric
\begin{equation}
ds^{2}=g_{tt}dt^{2}+g_{yy}dy^{2}+2g_{ty}dtdy+g_{rr}dr^{2}+d\overrightarrow{x}^{2},
\end{equation}
the orthogonal
metric is defined as
\begin{equation}
g_{\mu\nu}^{\bot}=(N_{t})_{\mu}(N_{t})_{\nu}+(N_{r})_{\mu}(N_{r})_{\nu},
\end{equation}
where $N_{t}$ and $N_{r}$ are unit normal vectors to the horizon
\begin{equation}
\begin{array}{ll}
N_{t}=\sqrt{\frac{g^{yy}}{g^{tt}g^{yy}-(g^{ty})^{2}}}(1,0,-\frac{g^{ty}}{g^{yy}},0),
\quad
N_{r}=(0,\frac{1}{\sqrt{g^{rr}}},0,0).\\
\end{array}
\end{equation}

After working out the components of the orthogonal metric and
Riemann tensor, for the metric (\ref{3eq1}) we can rewrite the
Wald formula (\ref{pp}) as
\begin{equation}
S_{BH}=\sum\limits_{i=1}^{4}S_{i} \label{3.5},
\end{equation}
where
\begin{eqnarray}
S_{1}&=&8\pi\int_{H}dx_{H}\sqrt{g_{H}}\frac{\partial\mathcal{L}}{\partial R_{trtr}}g_{tt}^{\bot}g_{rr}^{\bot}\nonumber\\
     &=&-32\pi Q_{5}\sqrt{Q_{2}Q_{6}}v_{1}\int_{H}dx_{H}\sqrt{g_{H}}\frac{\partial\mathcal{L}}{\partial R_{trtr}}R_{trtr},\nonumber\\
S_{2}&=&8\pi\int_{H}dx_{H}\sqrt{g_{H}}\frac{\partial\mathcal{L}}{\partial R_{yryr}}g_{yy}^{\bot}g_{rr}^{\bot}\nonumber\\
     &=&-\frac{32\pi Q_{5}Q_{P}^{2}\sqrt{Q_{2}Q_{6}}v_{1}}{r^{2}-Q_{P}^{2}}\int_{H}dx_{H}\sqrt{g_{H}}\frac{\partial\mathcal{L}}{\partial R_{yryr}}R_{yryr},\nonumber\\
S_{3}&=&16\pi\int_{H}dx_{H}\sqrt{g_{H}}\frac{\partial\mathcal{L}}{\partial R_{tryr}}g_{ty}^{\bot}g_{rr}^{\bot}\nonumber\\
     &=&-64\pi Q_{5}\sqrt{Q_{2}Q_{6}}v_{1}\int_{H}dx_{H}\sqrt{g_{H}}\frac{\partial\mathcal{L}}{\partial R_{tryr}}R_{tryr},\nonumber\\
S_{4}&=&8\pi\int_{H}dx_{H}\sqrt{g_{H}}\frac{\partial\mathcal{L}}{\partial R_{tyty}}(g_{tt}^{\bot}g_{yy}^{\bot}-(g_{ty}^{\bot})^{2})\nonumber\\
     &=&0.
\label{3.6}
\end{eqnarray}
 Next we define a function $f$ as integral of the Lagrangian density
over the horizon
\begin{equation}
f\equiv\int dx_{H}\sqrt{-\det g}\mathcal{L}.
\end{equation}
Following~\cite{sen}, we rescale the Riemann tensor components
\begin{equation}
\begin{array}{ll}
R_{rtrt}\rightarrow \lambda_{1}R_{rtrt}, \quad R_{ryry}\rightarrow
\lambda_{2}R_{ryry},\\
\\
R_{tryr}\rightarrow \lambda_{3}R_{tryr}, \quad
R_{tyty}\rightarrow\lambda_{4}R_{tyty}.\\
\\
\end{array}
\end{equation}
It can be seen that  the rescaled Lagrangian
$\mathcal{L}_{\lambda}$ behaves as
\begin{equation}
\frac{\partial
\mathcal{L}_{\lambda}}{\partial\lambda_{i}}=R_{\mu\nu\lambda\rho}^{(i)}\frac{\partial\mathcal{L}_{\lambda}}{\partial
R_{\mu\nu\lambda\rho}^{(i)}}.
\end{equation}
Note that there is no summation on the right hand side for $i$
($i=1,2,3,4$). Then we have the following relation for the rescaled
function $f_{\lambda}$
\begin{equation}
\frac{\partial
f_{\lambda}}{\partial\lambda_{i}}|_{\lambda_{i}=1}=v_{1}(Q_{P}+r)^{-\frac{1}{2}}Q_{5}^{\frac{1}{2}}\int
dx_{H}\sqrt{g_{H}}\frac{\partial\mathcal{L}_{\lambda}}{\partial
R_{\mu\nu\lambda\rho}^{(i)}}R_{\mu\nu\lambda\rho}^{(i)}.
\end{equation}
Substituting these  into (\ref{3.5}) and (\ref{3.6}), we find
\begin{equation}
S_{BH}=-8\pi\sqrt{Q_{2}Q_{5}Q_{6}(Q_{P}+r)}\left (\frac{\partial
f_{\lambda_{1}}}{\partial
\lambda_{1}}+\frac{Q_{P}^{2}}{Q_{P}^{2}-r^{2}}\frac{\partial
f_{\lambda_{2}}}{\partial \lambda_{2}}+\frac{\partial
f_{\lambda_{3}}}{\partial \lambda_{3}}\right )\left. \right
|_{\lambda_{1}=\lambda_{2}=\lambda_{3}=1}. \label{3.11}
\end{equation}

Since the general Lagrangian should be diffeomorphism invariant, the
components of the Riemann tensor appeared in (\ref{3.6}) must be
contracted by the corresponding components of the inverse metric.
Thus we have the following relations
\begin{eqnarray}
\lambda_{1}R_{trtr}g^{tt}g^{rr}\sim \lambda_{1}v_{1}^{-1},&\quad&\lambda_{2}R_{ryry}g^{rr}g^{yy}\sim
\lambda_{2}v_{1}^{-1},\nonumber\\
\lambda_{3}R_{tryr}g^{ty}g^{rr}\sim \lambda_{3}v_{1}^{-1},&\quad& \lambda_{4}R_{tyty}g^{tt}g^{yy}
\sim \lambda_{4}v_{1}^{-1}.
\end{eqnarray}
Furthermore, the electric field strength $F_{t49r}$ behaves as
$\sqrt{(g^{tt}g^{99}-(g^{t9})^{2})g^{rr}g^{99}}F_{t49r}\sim e_{1}v_{1}^{-\frac{3}{2}}$
and no other contributions have any dependence on $v_{1}$. These facts allow us to
rewrite $f_{\lambda}$ as a function of scalars, electric and magnetic field strengths
\begin{equation}
f_{\lambda}(u_{s},v_{1},v_{2},e_{1},p_{1},p_{2})=v_{1}^{\frac{3}{2}}h(u_{s},v_{2},\lambda_{i}
v_{1}^{-1},e_{1}v_{1}^{-\frac{3}{2}},p_{1},p_{2}),
\end{equation}
where $h$ is a general function and the factor
$v_{1}^{\frac{3}{2}}$ comes from $\sqrt{-\det g}$.

It can be easily derived that
\begin{equation}
\left. \sum\limits_{i=1}^{4}\lambda_{i}\frac{\partial
f_{\lambda_{i}}}{\partial\lambda_{i}}\right
|_{\lambda_{i}=1}=\frac{3}{2}(f-e_{1}\frac{\partial f}{\partial
e_{1}})-v_{1}\frac{\partial f}{\partial v_{1}}. \label{3.14}
\end{equation}
Then we can reexpress the entropy by substituting (\ref{3.14})
into (\ref{3.11})
\begin{equation}
S_{BH}=-8\pi\sqrt{Q_{2}Q_{5}Q_{6}(Q_{P}+r)}\{\frac{3}{2}(f-e_{1}\frac{\partial
f}{\partial e_{1}}) -(\frac{\partial
f}{\partial\lambda_{4}}-\frac{r^{2}}{Q_{P}^{2}-r^{2}}\frac{\partial
f}{\partial\lambda_{2}})\}.
\end{equation}
To simplify the above complicated expression, we have to make use
of the following relations, which are similar to those appeared
in~\cite{Ir}
\begin{equation}
\begin{array}{lll}
\frac{\partial f}{\partial \lambda_{1}}=\frac{\partial f}{\partial
\lambda_{2}},\quad 2\frac{\partial f}{\partial
\lambda_{1}}+\frac{\partial f}{\partial
\lambda_{3}}=2\frac{\partial f}{\partial \lambda_{4}},\quad
\frac{\partial f}{\partial \lambda_{3}}/\frac{\partial f}{\partial
\lambda_{2}}=\frac{2Q_{P}^{2}}{r^{2}-Q_{P}^{2}}.\\
\\
\end{array}
\end{equation}
Finally, the entropy has a simpler expression
\begin{eqnarray}
S_{BH}&=&-4\pi\sqrt{Q_{2}Q_{5}Q_{6}Q_{P}}(f-e_{1}\frac{\partial
f}{\partial
e_{1}})\nonumber\\
 &=&4\pi\sqrt{Q_{2}Q_{5}Q_{6}Q_{P}}F,
\end{eqnarray}
where $F\equiv e_{1}\frac{\partial f}{\partial e_{1}}-f$ and we
have used the fact that the horizon locates at $r=0$. Note that
although the parameters $v_{3}$ and $v_{4}$ appear in the deformed
near horizon metric, they do not contribute in the final form of
the entropy function because the (4,5,6,7,8) part is conformal to
a flat space metric. The entropy can be obtained by extremizing
the entropy function $F$ with respect to the moduli
\begin{equation}
\frac{\partial F}{\partial u_{s}}=0, \quad \frac{\partial
F}{\partial v_{i}}=0, \quad i=1,2.
\label{3.18}
\end{equation}
and then substituting the  values of the moduli back into $F$.

For the field configuration (\ref{3eq1}), we have
\begin{equation}
\mathcal{L}=\frac{1}{16\pi
G^{10}_{N}}[u_{s}(\frac{4v_{1}-3v_{2}}{2Q_{5}\sqrt{Q_{2}Q_{6}}v_{1}v_{2}}-\frac{1}{2v_{2}^{3}Q_{5}\sqrt{Q_{2}Q_{6}}})
-\frac{Q_{6}}{2v_{2}^{2}Q_{5}^{2}Q_{2}}+\frac{2Q_{2}Q_{6}Q_{p}^{2}}{Q_{5}^{2}v_{1}^{3}v_{2}}e_{1}^{2}].
\end{equation}
According to the definition of electric charge $
q_{1}=\frac{\partial f}{\partial e_{1}}$, we can obtain
\begin{equation}
e_{1}=\frac{v_{1}^{\frac{3}{2}}}{2Q_{2}Q_{P}v_{2}^{\frac{5}{2}}}.
\end{equation}
Then the function $f$ becomes
\begin{eqnarray}
f&=&\frac{(2\pi)^{2}R_{4}R_{9}V_{T^{4}}Q_{2}Q_{5}^{2}Q_{6}^{-1}v_{1}^{\frac{3}{2}}v_{2}^{\frac{7}{2}}}{4G^{10}_{N}}\nonumber\\
 & & \times [u_{s}(\frac{4v_{1}-3v_{2}}{2Q_{5}\sqrt{Q_{2}Q_{6}}v_{1}v_{2}}-\frac{1}{2Q_{5}\sqrt{Q_{2}Q_{6}}v_{2}^{3}})
-\frac{Q_{6}}{2Q_{2}Q_{5}^{2}v_{2}^{2}}+\frac{Q_{6}}{2Q_{2}Q_{5}^{2}v_{2}^{6}}].
\end{eqnarray}
and the entropy function
\begin{eqnarray}
F&=&\frac{(2\pi)^{2}R_{4}R_{9}V_{T^{4}}Q_{2}Q_{5}^{2}Q_{6}^{-1}v_{1}^{\frac{3}{2}}v_{2}^{\frac{7}{2}}}{4G^{10}_{N}}\nonumber\\
 & & \times [u_{s}(\frac{3v_{2}-4v_{1}}{2Q_{5}\sqrt{Q_{2}Q_{6}}v_{1}v_{2}}+\frac{1}{2Q_{5}\sqrt{Q_{2}Q_{6}}v_{2}^{3}})
+\frac{Q_{6}}{2Q_{2}Q_{5}^{2}v_{2}^{2}}+\frac{Q_{6}}{2Q_{2}Q_{5}^{2}v_{2}^{6}}].
\end{eqnarray}
The solutions to the moduli equations (\ref{3.18}) are
\begin{equation}
u_{s}=Q_{2}^{-\frac{1}{2}}Q_{5}^{-1}Q_{6}^{\frac{3}{2}}, \quad
v_{1}=v_{2}=1.
\end{equation}
Substituting these back to the entropy function, we arrive at
\begin{equation}
F=\frac{(2\pi)^{2}R_{4}R_{9}V_{T^{4}}}{4G^{10}_{N}},
\end{equation}
and the entropy of black hole is
\begin{eqnarray}
S_{BH}&=&4\pi\sqrt{Q_{2}Q_{5}Q_{6}Q_{P}}F\nonumber\\
      &=&2\pi\sqrt{N_{2}N_{5}N_{6}N_{P}}.
\end{eqnarray}
This is completely the same as the Bekenstein-Hawking entropy.

\section{Entropy function of 4-charge black hole in $D=4$}

In this section we  discuss the entropy function of 4-charge black
hole in four dimensions. For this end,  we first write down the
metric of the four dimensional extremal black hole in string frame
by dimensional reduction
\begin{equation}
\label{4eq1}
ds^{2(4)}_{str}=-(f_{2}f_{6})^{-\frac{1}{2}}(1+k)^{-1}dt^{2}
+(f_{2}f_{6})^{\frac{1}{2}}f_{s5}[dr^{2}+r^{2}(d\theta^{2}+\sin^{2}\theta
d\phi^{2})],
\end{equation}
where the functions $f_{2}, f_{s5}, f_{6}$ and $k$ remain the same
as before. To calculate the entropy function of this specific
background, we need the four dimensional effective action of type
IIA supergravity compactified on $S^{1}\times {S^{1}}'\times
T^{4}$. It can be obtained from the action (\ref{2eq1}) by the
standard dimensional reduction procedure (see, for example,
\cite{dr})
\begin{eqnarray}
\label{4eq2}
 S^{(4)}&=&\frac{1}{16\pi G_{N}^{4}}\int
d^{4}x\sqrt{-g^{(4)}}e^{\frac{\psi_{2}}{2}}
e^{\frac{\psi_{1}}{2}}e^{2\psi}\{e^{-2\phi}[R^{(4)}-\nabla^{2}\psi_{2}-\frac{1}{2}
(\nabla\psi_{2})^{2}-\frac{1}{4}e^{\psi_{2}}\mathcal{F}^{2}\nonumber\\
       & &-\nabla^{2}\psi_{1}-\frac{1}{2}(\nabla\psi_{1})^{2}+3(\nabla\psi)^{2}
       -8\nabla\psi\nabla\phi+4(\nabla\phi)^{2}-e^{-\psi_{1}}H^{(4)2}]\nonumber\\
       & &-G^{2}-e^{-\psi_{1}}e^{-\psi_{2}}F^{(4)2}\},
\end{eqnarray}
where $\psi$, $\psi_{1}$ and $\psi_{2}$ are single moduli for
$T^{4}$, ${S^{1}}'$ and $S^{1}$ respectively, $\phi$ is the ten
dimensional dilaton, $F^{(4)}$ is a two form electric field
strength coming from compactifying the ten dimensional four form
field strength and $\mathcal{F}$ is a two form field strength for
the gauge field corresponding to KK momentum along the $x_{9}$
direction. $H^{(4)}$ denotes the magnetic field strength coming
from compactifying the ten dimensional $(NS,NS)$ B field strength
and $G$ is another magnetic field strength coming from the ten
dimensional magnetic field strength associated with $D6$ branes,
which remains unchanged in the dimensional reduction.

For the black hole solution (\ref{4eq1}), the Wald formula can be
expressed as
\begin{equation}
S_{BH}=8\pi\int
dx_{H}\sqrt{g_{H}}\frac{\partial\mathcal{L}}{\partial
R_{trtr}}g_{tt}g_{rr}.
\end{equation}
Defining
\begin{equation}
f\equiv\int dx_{H}\sqrt{-g}\mathcal{L}
\end{equation}
and following the procedure used in the previous section, we have
\begin{eqnarray}
&& \left. \frac{\partial f}{\partial\lambda}\right
|_{\lambda=1}=4\int
dx_{H}\sqrt{-g}R_{trtr}\frac{\partial\mathcal{L}}{\partial
R_{trtr}}, \\
&& \left. \frac{\partial f}{\partial\lambda} \right
|_{\lambda=1}=f-e_{i}\frac{\partial f}{\partial e_{i}}, \\
&& S_{BH}=2\pi\sqrt{Q_{2}Q_{5}Q_{6}Q_{P}}(e_{i}\frac{\partial f}{\partial e_{i}}-f)
 \nonumber\\
      &&~~~~~~\equiv2\pi\sqrt{Q_{2}Q_{5}Q_{6}Q_{P}}F.
\end{eqnarray}

According to the black hole solution (\ref{4eq1}), we assume that
there exists a black hole configuration in the four dimensional
effective action (\ref{4eq2}) as follows,
\begin{eqnarray}
&ds_{str}^{2(4)}=v_{1}(-\frac{r^{2}}{Q_{P}\sqrt{Q_{2}Q_{6}}}dt^{2}+\frac{Q_{5}
\sqrt{Q_{2}Q_{6}}}{r^{2}}dr^{2})+v_{2}Q_{5}\sqrt{Q_{2}Q_{6}}(d\theta^{2}
+\sin^{2}\theta d\phi^{2}),\nonumber\\
&e^{\frac{\psi_{1}}{2}}=u_{1},\quad e^{\frac{\psi_{2}}{2}}=u_{2},
\quad e^{2{\psi}}=u_{T},\quad e^{-2{\phi}}=u_{s},\nonumber\\
&F^{(4)}_{rt}=e_{1},\quad \mathcal{F}_{rt}=e_{2},
 \quad H^{(4)}_{\theta\phi}=-\frac{Q_{5}}{2}\sin\theta, \quad G_{\theta\phi}=-\frac{Q_{6}}{2}\sin\theta.
\end{eqnarray}
Then we have
\begin{eqnarray}
f&=&\frac{1}{4G_{N}^{4}}[2u_{1}u_{2}u_{T}u_{s}(v_{1}-v_{2})Q_{5}
Q_{P}^{-\frac{1}{2}}-\frac{1}{2}v_{1}v_{2}^{-1}u_{1}^{-1}u_{2}u_{T}
u_{s}Q_{6}^{\frac{3}{2}}Q_{2}^{-\frac{1}{2}}Q_{5}^{-\frac{1}{2}}
Q_{P}^{-\frac{1}{2}}\nonumber\\
 & &-\frac{1}{2}v_{1}v_{2}^{-1}u_{1}u_{2}u_{T}Q_{5}^{\frac{3}{2}}
 Q_{2}^{-\frac{1}{2}}Q_{6}^{-\frac{1}{2}}Q_{P}^{-\frac{1}{2}}
 +2v_{1}^{-1}v_{2}u_{1}^{-1}u_{2}^{-1}u_{T}Q_{6}^{\frac{1}{2}}
 Q_{2}^{\frac{1}{2}}Q_{5}^{\frac{1}{2}}Q_{P}^{\frac{1}{2}}e_{1}^{2}\nonumber\\
 & &+\frac{1}{2}v_{1}^{-1}v_{2}u_{1}u_{2}^{3}u_{T}u_{s}Q_{6}^{\frac{1}{2}}
 Q_{2}^{\frac{1}{2}}Q_{5}^{\frac{1}{2}}Q_{P}^{\frac{1}{2}}e_{2}^{2}].
\label{4.9}
\end{eqnarray}
The parameters $e_{i}, i=1,2$ are related to the electric charges
$q_{i}$ via the equation $q_{i}=\partial f/\partial e_{i}$, so
that the values of $e_{i}$ can be determined as
\begin{eqnarray}
e_{1}&=&\frac{1}{2}v_{1}v_{2}^{-1}u_{1}u_{2}u_{T}^{-1}Q_{6}^{-\frac{1}{2}}
Q_{2}^{\frac{1}{2}}Q_{5}^{-\frac{1}{2}}Q_{P}^{-\frac{1}{2}},\nonumber\\
e_{2}&=&v_{1}v_{2}^{-1}u_{1}^{-1}u_{2}^{-3}u_{T}^{-1}u_{s}^{-1}
Q_{6}^{-\frac{1}{2}}Q_{2}^{-\frac{1}{2}}Q_{5}^{-\frac{1}{2}}Q_{P}^{\frac{1}{2}}.
\label{4.10}
\end{eqnarray}
Substituting (\ref{4.10}) into (\ref{4.9}), we obtain the entropy
function $F$
\begin{eqnarray}
F&=&\frac{1}{4G_{N}^{4}}[2u_{1}u_{2}u_{T}u_{s}(v_{2}-v_{1})Q_{5}
Q_{P}^{-\frac{1}{2}}+\frac{1}{2}v_{1}v_{2}^{-1}u_{1}^{-1}u_{2}
u_{T}u_{s}Q_{6}^{\frac{3}{2}}Q_{2}^{-\frac{1}{2}}Q_{5}^{-\frac{1}{2}}
Q_{P}^{-\frac{1}{2}}\nonumber\\
 & &+\frac{1}{2}v_{1}v_{2}^{-1}u_{1}u_{2}u_{T}Q_{5}^{\frac{3}{2}}
 Q_{2}^{-\frac{1}{2}}Q_{6}^{-\frac{1}{2}}Q_{P}^{-\frac{1}{2}}
 +\frac{1}{2}v_{1}v_{2}^{-1}u_{1}u_{2}u_{T}^{-1}Q_{6}^{-\frac{1}{2}}
 Q_{2}^{\frac{3}{2}}Q_{5}^{-\frac{1}{2}}Q_{P}^{-\frac{1}{2}}\nonumber\\
 & &+\frac{1}{2}v_{1}v_{2}^{-1}u_{1}^{-1}u_{2}^{-3}u_{T}^{-1}
 u_{s}^{-1}Q_{6}^{-\frac{1}{2}}Q_{2}^{-\frac{1}{2}}Q_{5}^{-\frac{1}{2}}Q_{P}^{\frac{3}{2}}].
\end{eqnarray}
The black hole entropy can be obtained by extremizing the entropy
function $F$ with respect to the moduli,
\begin{equation}
\frac{\partial F}{\partial v_{i}}=0,\quad i=1,2 \quad\quad
\frac{\partial F}{\partial u_{i}}=0,\quad i=1,2,T,s,
\end{equation}
from which we  have
\begin{equation}
\begin{array}{lll}
v_{1}=v_{2}=v,\\
\\
u_{1}=Q_{6}^{\frac{3}{4}}Q_{2}^{-\frac{1}{4}}Q_{5}^{-\frac{3}{4}}v^{-\frac{1}{2}},
\quad
u_{2}=Q_{6}^{-\frac{1}{4}}Q_{2}^{-\frac{1}{4}}Q_{5}^{\frac{1}{4}}Q_{P}^{\frac{1}{2}}v^{\frac{1}{2}},\\
\\
u_{T}=\frac{Q_{2}}{Q_{5}},\quad
u_{s}=Q_{6}^{-\frac{1}{2}}Q_{2}^{-\frac{1}{2}}Q_{5}^{\frac{1}{2}}v^{-1}.\\
\\
\end{array}
\end{equation}
Here $v$ is an arbitrary constant, which will not appear in the
entropy of black hole.  With these results, the entropy function
reduces to
\begin{equation}
F=\frac{2}{4G_{N}^{4}}.
\end{equation}
The black hole entropy then becomes
\begin{equation}
S_{BH}=2\pi\sqrt{N_{2}N_{5}N_{6}N_{P}}.
\end{equation}
Clearly we have obtained the black hole entropy once again.

\section{Higher-order corrections to entropy of 4-charge black hole}

In this section we will compute the corrections to the entropy
function by making use of the low-energy effective action for type
IIA superstrings with $\alpha'^3$
corrections~\cite{cor}~\cite{ts}~\cite{bb}~\cite{ohta}. The
corrections in string frame can be written as~\cite{ohta}
\begin{equation}
S^{(cor)}_{IIA}=\frac{1}{16\pi G^{10}_{N}}\int d^{10}x \sqrt{-g}
{\alpha'}^{3}[\frac{\pi^{2}}{3^{2}\cdot
2^{8}}\widetilde{E}_{8}+(\frac{\zeta(3)}{2^{3}}e^{-\frac{\phi}{2}}+\frac{\pi^{2}}{3\cdot
2^{3}}e^{\frac{3\phi}{2}})L_{W}],
\end{equation}
where
\begin{eqnarray}
\widetilde{E}_{2n}&=&-\frac{1}{2^{n}(D-2n)!}\epsilon^{\alpha_{1}\cdots
\alpha_{D-2n}\mu_{1}\nu_{1}\cdots
\mu_{n}\nu_{n}}\epsilon_{\alpha_{1}\cdots
\alpha_{D-2n}\rho_{1}\sigma_{1}\cdots
\rho_{n}\sigma_{n}}{R^{\rho_{1}\sigma_{1}}}_{\mu_{1}\nu_{1}}\cdots
{R^{\rho_{n}\sigma_{n}}}_{\mu_{n}\nu_{n}},\nonumber\\
L_{W}&=&C^{\lambda\mu\nu\kappa}C_{\alpha\mu\nu\beta}{C_{\lambda}}^{\rho\sigma\alpha}{C^{\beta}}_{\rho\sigma\kappa}
+\frac{1}{2}C^{\lambda\kappa\mu\nu}C_{\alpha\beta\mu\nu}{C_{\lambda}}^{\rho\sigma\alpha}{C^{\beta}}_{\rho\sigma\kappa}.
\end{eqnarray}
Assuming that the near horizon geometry still has the form
(\ref{3eq1}) when the higher derivative terms are taken into
account, and following the same steps, we get the corrected
entropy function
\begin{eqnarray}
F&=&\frac{(2\pi)^{2}R_{4}R_{9}V_{T^{4}}Q_{2}Q_{5}^{2}Q_{6}^{-1}v_{1}^{\frac{3}{2}}v_{2}^{\frac{7}{2}}}{4G^{10}_{N}}
[u_{s}(\frac{3v_{2}-4v_{1}}{2Q_{5}(Q_{2}Q_{6})^{\frac{1}{2}}v_{1}v_{2}}+\frac{1}{2Q_{5}
(Q_{2}Q_{6})^{\frac{1}{2}}v_{2}^{3}})+\frac{Q_{6}}{2Q_{2}Q_{5}^{2}v_{2}^{2}}\nonumber\\
 & &
+\frac{Q_{6}}{2Q_{2}Q_{5}^{2}v_{2}^{6}}-5{\alpha'}^{3}(p_{1}{u_{s}}^{\frac{1}{4}}+p_{2}
{u_{s}}^{-\frac{3}{4}})\frac{(14336v_{1}^{4}-1536v_{1}^{3}v_{2}+1728v_{1}^{2}v_{2}^{2}
-432v_{1}v_{2}^{3}+567v_{2}^{4})}{222184Q_{2}^{2}Q_{5}^{4}Q_{6}^{2}v_{1}^{4}v_{2}^{4}}],
\end{eqnarray}
where $p_{1}\equiv\frac{\zeta(3)}{2^{3}}$ and
$p_{2}\equiv\frac{\pi^{2}}{3\cdot 2^{3}}$.  Note that for the
configuration (\ref{3eq1}), the $\widetilde{E}_{8}$ term does not
have any contribution here. In addition,
  it is interesting to
see that the parameters $v_{3}$ and $v_{4}$ still do not appear in
the entropy function.

The solutions to the equations of motion of moduli by extremizing
the entropy function are found to be
\begin{eqnarray}
v_{1}&=&1+{{\alpha'}}^{3}\frac{747813p_{2}Q_{2}Q_{5}\sqrt{Q_{6}}-249271p_{1}\sqrt{Q_{2}}{Q_{6}}^{2}+7712
p_{2}Q_{2}Q_{5}\sqrt{Q_{6}}+7712p_{1}\sqrt{Q_{2}}{Q_{6}}^{2}}
{3554944{Q_{5}}^{\frac{9}{4}}{Q_{2}}^{\frac{13}{8}}{Q_{6}^{\frac{37}{8}}}},\nonumber\\
v_{2}&=&1+{{\alpha'}}^{3}\frac{23136p_{2}Q_{2}Q_{5}\sqrt{Q_{6}}+23136p_{1}\sqrt{Q_{2}}{Q_{6}}^{2}+483879
p_{2}Q_{2}Q_{5}\sqrt{Q_{6}}-161293p_{1}\sqrt{Q_{2}}{Q_{6}}^{2}}
{3554944{Q_{5}}^{\frac{9}{4}}{Q_{2}}^{\frac{13}{8}}{Q_{6}^{\frac{37}{8}}}},\nonumber\\
u_{s}&=&\frac{Q_{6}^{\frac{3}{2}}}{Q_{5}Q_{2}^{\frac{1}{2}}}(1-{{\alpha'}}^{3}\frac{2067483p_{2}Q_{2}Q_{5}\sqrt{Q_{6}}-689161p_{1}\sqrt{Q_{2}}{Q_{6}}^{2}+1103632
p_{2}Q_{2}Q_{5}\sqrt{Q_{6}}+1103632p_{1}\sqrt{Q_{2}}{Q_{6}}^{2}}{3554944{Q_{5}}^{\frac{5}{4}}
{Q_{2}}^{\frac{9}{8}}{Q_{6}^{\frac{49}{8}}}}).
\end{eqnarray}
Finally we obtain the corrected entropy of black hole
\begin{equation}
S=2\pi\sqrt{N_{2}N_{5}N_{6}N_{p}}[1+{\alpha'}^{3}\frac{3(39355p_{2}\sqrt{Q_{2}}Q_{5}-136601p_{1}
Q_{6}^{\frac{3}{2}})}{888736Q_{2}^{\frac{9}{8}}Q_{5}^{\frac{9}{4}}Q_{6}^{\frac{33}{8}}}].
\end{equation}

However, if we use the higher derivative corrections in~\cite{Ir}
\begin{equation}
\begin{array}{ll}
L_{corr}=\gamma e^{-2\phi}(L_{1}-2L_{2}+\lambda L_{3}), \\
\\
L_{1}=R^{hmnk}R_{pmnq}{R_{h}}^{rsp}{R^{q}}_{rsk}+\frac{1}{2}R^{hkmn}R_{pqmn}{R_{h}}^{rsp}{R^{q}}_{rsk},\\
\\
L_{2}=R^{hk}(\frac{1}{2}R_{hnpk}R^{msqn}{R_{msq}}^{p}+\frac{1}{4}R_{hpmn}{R_{k}}^{pqs}{R_{qs}}^{mn}+R_{hmnp}{R_{kqs}}^{p}R^{nqsm}),\\
\\
L_{3}=R(\frac{1}{4}R_{hpmn}R^{hpqs}{R_{qs}}^{mn}+R_{hmnp}{{R^{h}}_{qs}}^{p}R^{nqsm}),\\
\end{array}
\end{equation}
where $\gamma=\frac{1}{8}\zeta(3)\alpha'^{3}$ and $\lambda$ is a
parameter which signifies the ambiguity in the field redefinitions
of the metric, we will obtain different results. To see this, we
follow similar steps mentioned above and arrive at the following
entropy function
\begin{eqnarray}
F&=&\frac{(2\pi)^{2}R_{4}R_{9}V_{T^{4}}Q_{2}Q_{5}^{2}Q_{6}^{-1}v_{1}^{\frac{3}{2}}
v_{2}^{\frac{7}{2}}}{4G^{10}_{N}}\{u_{s}(\frac{3v_{2}-4v_{1}}{2Q_{5}\sqrt{Q_{2}Q_{6}}v_{1}v_{2}}
+\frac{1}{2Q_{5}\sqrt{Q_{2}Q_{6}}v_{2}^{3}})\nonumber\\
 & &+\frac{Q_{6}}{2Q_{2}Q_{5}^{2}v_{2}^{2}}+\frac{Q_{6}}{2Q_{2}Q_{5}^{2}v_{2}^{6}}
 +\gamma u_{s}[\frac{4}{Q_{5}^{4}{(Q_{2}Q_{6})}^{2}v_{2}^{4}}-\frac{7\cdot3^{4}}
 {2^{4}6^{3}Q_{5}^{4}{(Q_{2}Q_{6})}^{2}v_{1}^{4}}\nonumber\\
 & &-\lambda\frac{4v_{2}-3v_{1}}{2Q_{5}{(Q_{2}Q_{6})}^{\frac{1}{2}}v_{1}v_{2}}
 (\frac{4}{Q_{5}^{3}{(Q_{2}Q_{6})}^{\frac{3}{2}}v_{2}^{3}}-\frac{3^{3}}{2^{3}
 6^{2}Q_{5}^{3}{(Q_{2}Q_{6})}^{\frac{3}{2}}v_{1}^{3}})]\}.
\end{eqnarray}
The solutions to the corresponding moduli equations are
\begin{eqnarray}
&& v_{1}=1+\frac{4363-3358\lambda}{1600}\frac{\gamma}{Q_{5}^{3}(Q_{2}Q_{6})^{\frac{3}{2}}}, \nonumber \\
&& v_{2}=1+\frac{814-3524\lambda}{1600}\frac{\gamma}{Q_{5}^{3}(Q_{2}Q_{6})^{\frac{3}{2}}},\nonumber \\
&&
u_{s}=\frac{Q_{6}^{\frac{3}{2}}}{Q_{5}Q_{2}^{\frac{1}{2}}}(1+\frac{407+4038\lambda}{1600}\frac{\gamma}{Q_{5}^{3}(Q_{2}Q_{6})^{\frac{3}{2}}}).
\end{eqnarray}
The corrected entropy becomes
\begin{equation}
S=2\pi\sqrt{N_{2}N_{5}N_{6}N_{p}}(1+\frac{491-262\lambda}{128}\frac{\gamma}{N_{5}^{3}(N_{2}N_{6})^{\frac{3}{2}}}(\frac{2R_{4}}{G^{4}_{N}\alpha'R_{9}})^{\frac{3}{2}}).
\end{equation}
Note that the corrected entropy depends on the coefficient
$\lambda$, which is different from the 3-charge solution case in
type IIB supergravity~\cite{Ir}. Because in~\cite{Ir} the $AdS_{3}$
part and $S_{3}$ part have the same curvature radii so that the term
related to $\lambda$ does not contribute, while here the two parts
$AdS_3$ and $S^2$ have different radii and the $\lambda$ term
appeared in the final result.

As pointed out in~\cite{cor}, ~\cite{ts} and~\cite{bb}, the field
redefinition ambiguity allows one to choose the corrected action in
a specific ''scheme" where only the Weyl tensor part of the
curvature appears in the action. If we consider the Weyl tensor part
only, i.e. the corrections to the effective action turn out to be
\begin{equation}
L_{corr}=\gamma
e^{-2\phi}(C^{hmnk}C_{pmnq}{C_{h}}^{rsp}{C^{q}}_{rsk}+\frac{1}{2}C^{hkmn}C_{pqmn}{C_{h}}^{rsp}{C^{q}}_{rsk}),
\end{equation}
then the corrected entropy function is
\begin{eqnarray}
F&=&\frac{(2\pi)^{2}R_{4}R_{9}V_{T^{4}}Q_{2}Q_{5}^{2}Q_{6}^{-1}v_{1}^{\frac{3}{2}}v_{2}^{\frac{7}{2}}}{4G^{10}_{N}}[u_{s}(\frac{3v_{2}-4v_{1}}{2Q_{5}(Q_{2}Q_{6})^{\frac{1}{2}}v_{1}v_{2}}+\frac{1}{2Q_{5}(Q_{2}Q_{6})^{\frac{1}{2}}v_{2}^{3}})\nonumber\\
 & &+\frac{Q_{6}}{2Q_{2}Q_{5}^{2}v_{2}^{2}}+\frac{Q_{6}}{2Q_{2}Q_{5}^{2}v_{2}^{6}}-\gamma u_{s}\frac{5(14336v_{1}^{4}-1536v_{1}^{3}v_{2}+1728v_{1}^{2}v_{2}^{2}-432v_{1}v_{2}^{3}+567v_{2}^{4})}{222184Q_{2}^{2}Q_{5}^{4}Q_{6}^{2}v_{1}^{4}v_{2}^{4}}].
\end{eqnarray}
In this case,  the solutions to the corresponding moduli equations
become
\begin{eqnarray}
&& v_{1}=1-\frac{2278857}{349495432}\frac{\gamma}{Q_{5}^{3}(Q_{2}Q_{6})^{\frac{3}{2}}}, \nonumber \\
&& v_{2}=1+\frac{775009}{31772312}\frac{\gamma}{Q_{5}^{3}(Q_{2}Q_{6})^{\frac{3}{2}}}, \nonumber \\
&&
u_{s}=\frac{Q_{6}^{\frac{3}{2}}}{Q_{5}Q_{2}^{\frac{1}{2}}}(1-\frac{128859193}{349495432}\frac{\gamma}{Q_{5}^{3}(Q_{2}Q_{6})^{\frac{3}{2}}}).
\end{eqnarray}
And the corrected entropy is
\begin{equation}
S=2\pi\sqrt{N_{2}N_{5}N_{6}N_{p}}[1-\frac{73315}{222184}\frac{\gamma}{N_{5}^{3}(N_{2}N_{6})^{\frac{3}{2}}}
(\frac{2R_{4}}{G^{4}_{N}\alpha'R_{9}})^{\frac{3}{2}}].
\end{equation}
From the above we can draw the conclusion that the corrected
entropy depends on the exact form of the corrected action in the
case of  4-charge black holes in type IIA supergravity.

\section{Conclusion}

Sen's entropy function method turns out to be a useful tool for
calculating the entropy of extremal black holes whose near horizon
geometry are $AdS_{2}\times S^{D-2}$. In this paper we have
calculated the entropy function of 4-charge extremal black holes in
type IIA supergravity both in 10 and 4 dimensions and found that the
resulting entropy of the black hole is in precise agreement with the
Bekenstein-Hawking entropy. Note that the near-horizon geometry of
the black hole in 10 dimensions does not have the form $AdS_2 \times
S^{D-2}$, instead it is $AdS_3 \times S^2 \times S^1 \times T^4$.
Combining the result for 3-charge black hole considered in
\cite{Ir}, where the black hole geometry is $AdS_3 \times S^3 \times
T^4$, we can see that the Sen's entropy function method is not
limited to the geometry $AdS_{2}\times S^{D-2}$ only.

We have found that there exist some ambiguities in calculating the
higher order corrections to the entropy. First, we have computed the
corrections by making use of the $\alpha'^3$ corrections in the
tree-level and one-loop action. Then we have found that when taking
the corrected action used in~\cite{Ir}, the parameter $\lambda$,
which stands for the ambiguity of the field redefinition, appears in
the final result. This curious phenomenon arises due to the fact
that the $AdS_{3}$ part and $S^{2}$ part have different curvature
radii. This is quite different from the 3-charge case considered
in~\cite{Ir}. Furthermore, the corrections are quite different if we
work them out using various forms of the action. It means that the
corrections to the entropy depend on the exact form of the action.
In other words, the corrections depends on the schemes. It would be
very important to further understand the dependence of the schemes.

We notice that the field redefinition ambiguities and scheme
dependence for black hole entropy in heterotic string theory has
been discussed extensively in~\cite{Ex1}, where the author pointed
out that since Wald's formula was applied on the local horizon which
was the exact solution of the truncated equations of motion then
Wald's entropy depended on the field redefinition ambiguity
parameters. However, this problem can be solved by requiring that
the result obtained via the entropy function formalism should agree
with the statistical entropy. It has been found that there exist
schemes in which the inclusion of all the linear $\alpha'$
corrections gives rise to a `local' horizon with geometry
$AdS_{2}\times S^{D-2}$ and for which the modified
Bekenstein-Hawking entropy is in agreement with the statistical
entropy. So it seems that requiring the agreement between
macroscopic entropy and microscopic entropy is an effective method
for removing the scheme dependence. Similarly, the scheme dependence
also exists in our case and we expect that it can also be resolved
likely. However, we still do not have a clear idea about how to
choose the proper scheme following~\cite{Ex1} due to the lack of
understanding the corrections to the entropy from a microscopic
point of view. We hope to address this problem in our future work.

Finally, we would like to mention that we still assume the geometry
of black hole being of the form (\ref{3eq1}) when the higher
derivative terms are taken into account.  This should be justified.
However, it is a difficult problem. Here we give some arguments to
support this assumption. One piece of evidence is that under such an
assumption, we have obtained nontrivial and self consistent
solutions to the moduli equations. Furthermore, let us consider the
following action
\begin{align} S&=\frac{1}{2\kappa_D^2}\int d^Dx\,\sqrt{-g}\left( R
+\alpha_4E_8
 +\gamma J_0\right)
\end{align}
where $E_8, J_0$ are given as
\begin{align}
E_8&=-\frac{1}{2^4\times(D-8)!}\epsilon^{\alpha_1\alpha_2
\cdots\alpha_{D-8} \rho_1\sigma_1
\cdots\rho_4\sigma_4}\epsilon_{\alpha_1\alpha_2\cdots
\alpha_{D-8}\mu_1\nu_1\cdots\mu_4\nu_4}R^{\mu_1\nu_1}\!_{\rho_1\sigma_1}
 R^{\mu_2\nu_2}\!_{\rho_2\sigma_2} R^{\mu_3\nu_3}\!_{\rho_3\sigma_3}
 R^{\mu_4\nu_4}\!_{\rho_4\sigma_4} \\
J_0&=C^{\lambda\mu\nu\kappa}C_{\alpha\mu\nu\beta}C^{\;\;\alpha\rho\sigma}_{\lambda}
C^{\beta}_{\;\rho\sigma\kappa}+\frac{1}{2}C^{\lambda\kappa\mu\nu}
C_{\alpha\beta\mu\nu}C_{\lambda}^{\;\;\rho\sigma\alpha}
C^{\beta}_{\;\rho\sigma\kappa}
\end{align}
where $C$ is the Weyl tensor. Then
\begin{eqnarray}
ds^2_D=-f dt^2+f^{-1}dr^2+r^2 h_{ij} dx^i dx^j
\end{eqnarray}
where $h_{ij}$ is the metric for maximally symmetric space of
$D-2$ dimensions. Taking it to be sphere, we find that
$f=1+(r/l)^2$ is an exact solution of the system with $l^6 =
\alpha_4 (D-3)(D-4) \cdots (D-8)$. Note that here $\gamma$ term
does not contribute. In our case, if we do dimensional reduction
in (\ref{3eq1}) on $S^2\times S^1 \times T^4$, only the $AdS_3$ is
left. Taking into account the higher derivative terms, we can see
that the $AdS_3$ is still an exact solution; only difference is to
change the radius of $AdS_3$ space, which correspondingly changes
the size $v_1$.  Of course, it would be very interesting to
rigorously prove the assumption.

\section*{Acknowledgements}

DWP thanks Jian-Wei Cui for doing the calculations on his computer
and Hua Bai, Li-Ming Cao, Qing-Guo Huang, Hui Li, Xun Su and
Wei-Shui Xu for useful discussions and kind help.  We would like
to thank N. Ohta for a careful reading and useful suggestions and
discussions. This work was supported by a grant from Chinese
Academy of Sciences, and grants from NSFC, China (No. 10325525 and
No.90403029).

\end{document}